\documentclass[prb,showpacs,aps,floats,floatfix,superscriptaddress]{revtex4}

\usepackage{epsfig}

\begin{document}

\title{\bf Dispersion Interactions between Optically Anisotropic Cylinders at all Separations: \\
Retardation Effects for Insulating and Semiconducting Single Wall Carbon Nanotubes} 

% RUDI, I changed the title here. As far as I understand, this theory cannot work for metallic SWCNTs.

\author{A. \v Siber $\diamond$, R.F. Rajter $\dagger\dagger$, R.H. French $\S$, W.Y. Ching$\S\dagger$, V.A. Parsegian$\dagger$ and R. Podgornik $\dagger$$\star$
\\
 [3mm] 
 $\diamond$ Institute of Physics, Bijeni\v{c}ka cesta 46, P.O. Box 304, 10001 Zagreb, Croatia \\
 $\dagger$ Laboratory of Physical and Structural Biology, NICHD, Bld. 9, Rm. 1E116 \\
 National Institutes of Health, Bethesda, MD 20892-0924, USA\\
 $\dagger\dagger$ Department of Materials Science and Engineering, MIT, Rm 13-5046, USA\\ 
 $\star$ Dept. of Physics, Faculty of Mathematics and Physics and Institute of Biophysics, School of Medicine, University of Ljubljana, Ljubljana, Slovenia and Dept. of Theoretical Physics, J. Stefan Institute, Ljubljana, Slovenia\\
 \S DuPont Co., Central Research, E400-5407, Wilmington DE 19880, USA \\
 $\S\dagger$Department of Physics, University of Missouri-Kansas
City Kansas City, Missouri, 64110, USA }

%\textwidth 16 true cm
%\textheight 22 true cm
%\hoffset=-15 mm
%\voffset=-10 mm
%\begin{document}
    
\begin{abstract}
We derive the complete form of the van der Waals dispersion interaction between two infinitely long anisotropic semiconducting/insulating thin cylinders at all separations. The derivation is based on the general theory of dispersion interactions between anisotropic media as formulated in  [J. N. Munday, D. Iannuzzi, Yu. S. Barash and F. Capasso, {\sl Phys. Rev. A} {\bf 71}, 042102 (2005)]. This formulation is then used to calculate the dispersion interactions between a pair of single walled carbon nanotubes at all separations and all angles. Non-retarded and retarded forms of the interactions are developed separately. The possibility of repulsive dispersion interactions and non-monotonic dispersion interactions is discussed within the framework of the new formulation.
\end{abstract}


\maketitle        

\section{Introduction}

Single walled carbon nanotubes (SWCNTs) constitute a unique class of materials with chirality-dependent dielectric properties \cite{dressel} that give rise to interesting consequences in terms of their van der Waals - dispersion  interactions. Several experimental procedures have exploited the differences among these properties in order to separate SWCNTs by chirality (see Ref. \onlinecite{rick1} and references therein). In order to reliably separate a polydisperse solution of SWCNTs into monodisperse fractions of one chirality, one needs to understand the detailed features of the interactions between constituent SWCNTs. Apart from this, the dispersion interactions could also drive micro and nano-mechanical actuators that could transduce rectilinear motion or even convert rectilinear into rotational motion \cite{capasso-IEEE}.  To achieve this goal a much more rigorous understanding of  fundamental forces and, in particular, van der Waals -  dispersion interactions between SWCNTs is needed. Several important advances have already been made \cite{rick2}. Our work continues these efforts.

We derive the complete van der Waals -  dispersion interaction free energy between two anisotropic cylinders at all separations, including the retardation effects. We start with the Lifshitz theory of van der Waals interactions between two semiinifinte anisotropic uniaxial dielectric layers across a finite layer of dielectric function $\epsilon_{m}$ and thickness $\ell$ as worked out by Barash \cite{Barash}, the result of this calculation is the interaction free energy between the two layers as a function of their separation $\ell$ and the angle between their principal dielectric anisotropy axes $\theta$: ${\cal G}(\ell,\theta)$. The dielectric response of the two dielectrically uniaxial half-spaces is given by the values of their dielectric functions $\overline{\epsilon_{\parallel}}$, parallel and $\overline{\epsilon_{\perp}}$, perpendicular to their respective axes.  We shall use $\overline{\epsilon_{1,\parallel}}$ ($\overline{\epsilon_{1,\perp}}$) and $\overline{\epsilon_{2,\parallel}}$ ($\overline{\epsilon_{2,\perp}}$) for the left and right half-spaces, respectively. Note also that in the theory of van der Waals interactions \cite{Parsegian} all the dielectric response functions are evaluated at imaginary frequencies, thus $\epsilon_{\parallel,\perp} = \epsilon_{\parallel,\perp}(i \omega)$. $\epsilon_{\parallel,\perp}(i \omega)$ is referred to as the London - van der Waals transform of the response function $\epsilon_{\parallel,\perp}(\omega)$ and is given by the Kramers - Kronig relations. It is strictly a real, monotonically decaying function of $\omega$. 

From the interaction free energy between two half-spaces one can extract the interaction between two cylinders by assuming that the two half-spaces are 
dilute assemblies of anisotropic cylinders. This derivation closely follows the arguments of Pitaevskii for evaluating the interactions between isotropic impurity 
atoms in a homogeneous fluid \cite{Pitaevskii}. 
We assume that the two anisotropic half-spaces are composed of anisotropic cylinders at volume fractions $v_1$ and $v_2$, with 
${\epsilon^{c}}_{1,\perp}$ (${\epsilon^{c}}_{2,\perp}$) and ${\epsilon^{c}}_{1,\parallel}$ (${\epsilon^{c}}_{2,\parallel}$) as the 
transverse and longitudinal dielectric response functions of the cylinder materials. We then expand ${\cal G}(\ell,\theta)$ for two half-spaces 
as a series in $v_1$ and $v_2$ and evaluate the coefficient multiplying the $v_1 v_2$ term. The volume 
fractions $v_1$ and $v_2$ scale with the area density of the cylinders ($N_1, N_2$) in the direction of their long axes as $v_1 = N_1~\pi R_1^{2}$ ($v_2 = N_2~\pi R_2^{2}$). It then follows \cite{Parsegian} that the interaction free energy between two cylinders, $G(\ell,\theta)$, whose axes are contained within the two parallel boundaries 
at a separation $\ell$, but skewed at an angle 
$\theta$ is given by 
\begin{equation}
\frac{d^{2}{\cal G}(\ell,\theta)}{d\ell^{2}}= N_1 N_2\sin{\theta}~G(\ell,\theta).
\label{form1}
\end{equation}
Conversely, the interaction free energy {\sl per unit length}, $g(\ell)$, between two parallel cylinders is given by the Abel transform
\begin{equation}
\frac{d^{2}{\cal G}(\ell,\theta=0)}{d\ell^{2}}=N_1 N_2\int_{-\infty}^{+\infty}g(\sqrt{\ell^{2}+y^{2}})~dy.
\label{form2}
\end{equation}
In both cases we  expand ${\cal G}(\ell,\theta)$ to find the coefficient next to $v_1 v_2$ (or equivalently $N_1 N_2$), take the second derivative with respect to $\ell$, then use Eqs. \ref{form1} and \ref{form2} in order to obtain the appropriate pair interaction free energy between cylinders. Note that such an expansion is possible only if the dielectric response at all frequencies is bounded. In the case of an ideal metal Drude-like dielectric response this expansion is not feasible and our method can not be transplanted to that case automatically.

The closest attempt in the literature to evaluate the interaction between two cylinders at all separations comes from Barash and Kyasov \cite{Barash89}. 
Where this approach can be compared with the one presented here, {\sl i.e.} for two parallel isotropic cylinders, the results for the interaction free energy 
between parallel cylinders coincide completely. As far as we are aware our calculation thus presents the first attempt to evaluate the van der Waals - dispersion interaction 
between two skewed anisotropic cylinders at all angles and at all separations. Its major drawback is that by construction it is valid only for materials with finite dielectric 
response and thus cylinders with ideally metallic dielectric functions showing a Drude-like peak at zero frequency can not be treated by the theory presented below. 
Also the calculation is only valid for infinitely long cylinders and finite length effect are not taken into account exactly.

\section{Derivation}

We use the Pitaevskii {\sl ansatz} in order to extract the interactions between two infinite anisotropic cylinders at all separations and angles from the interaction between 
two semi-infinite half-spaces of anisotropic uniaxial dielectric material. We start with the fully retarded van der Waals - dispersion interactions between two semiinfinite anisotropic 
dielectric slabs \cite{Barash}. The full interaction form is quite involved, but it has a simple limit if the two semiinfinite slabs, ${\cal R}$ and ${\cal L}$, separated by an isotropic medium of thickness $\ell$, are composed of rarefied material. 

In order to get the interaction free energy between two anisotropic cylinders we assume that both semi-infinite substrates (half-spaces), ${\cal L}$ ($1$) and ${\cal R}$ ($2$), are 
composite materials made of oriented anisotropic cylinders at volume fractions $v_1$ and $v_2$, with ${\epsilon^{c}}_{1,\perp}$ (${\epsilon^{c}}_{2,\perp}$) 
and ${\epsilon^{c}}_{1,\parallel}$ (${\epsilon^{c}}_{2,\parallel}$) as the transverse and longitudinal
dielectric response functions of the cylinder materials. For the semi-infinite composite medium of oriented anisotropic cylinders with local hexagonal  packing symmetry, so that the corresponding cylinder volume fraction is $v$, the anisotropic bulk dielectric response function can be derived in the
form (see Ref. \onlinecite{Parsegian}, p.318) 
\begin{equation}
\overline{\epsilon_{\parallel}}=\epsilon_{m}\left(1+v\Delta_{\parallel}\right),\qquad\overline{\epsilon_{\perp}}=\epsilon_{m}\left(1+\frac{2v\Delta_{\perp}}{1-v\Delta_{\perp}}\right),\label{eq:v_dependance}
\end{equation}
where the relative anisotropy measures in the parallel and perpendicular direction are given by
\begin{equation}
\Delta_{\perp}=\frac{{\epsilon^{c}}_{\perp}-\epsilon_{m}}{{\epsilon^{c}}_{\perp}+\epsilon_{m}}\qquad\Delta_{\parallel}=\frac{{\epsilon^{c}}_{\parallel}-\epsilon_{m}}{\epsilon_{m}}.
\label{anisoind}
\end{equation}
In our case, this holds for both ${\cal L}$ and ${\cal R}$ half-spaces with the appropriate volume fractions and dielectric responses. $\epsilon_{m}$ is the dielectric function of the 
isotropic medium between the cylinders as well as between regions ${\cal L}$ and ${\cal R}$. We assume in what follows that all the response functions are bounded and finite.
 
The formulae in Eqs. \ref{form1},\ref{form2} connect the interaction free energy of two semiinifinite half spaces with the interaction free energy between two cylinders either 
parallel or skewed at a finite angle $\theta$. The Barash result \cite{Barash} for the complete retarded form of the interactions between two uniaxial media, ${\cal G}(\ell,\theta)$, 
is quite complicated (note also a typo that propagated starting from the original version of the calculation \cite{erratum}) but can be straightforwardly expanded to second order 
in $N$ (a term proportional to $v_1 v_2$) for the dielectric response functions of the form Eq. \ref{eq:v_dependance}, yielding the following result
\begin{equation}
\frac{d^{2}{\cal G}(\ell,\theta)}{d\ell^{2}} = \frac{k_BT}{2\pi} {\sum_{n=0}^{\infty}}' \int_0^{\infty} Q dQ \frac{d^{2}f(\ell,\theta)}{d\ell^{2}}.
\end{equation}
In the above equation, $n$ represent the (thermal) Matsubara indices, the prime on the summation means  that the weight of the $n=0$ term is 1/2 (see Refs. \onlinecite{Parsegian,Barash89} for details). The second derivative of the function $f(\ell,\theta)$ can be obtained explicitly 
in terms of the ratios between the relative anisotropy measures (Eq. \ref{anisoind}) defined as 
\begin{equation}
a = \frac{2 \Delta_{\perp}}{\Delta_{\parallel}} = 2 \frac{({\epsilon^{c}}_{\perp}-\epsilon_{m}) \epsilon_{m}}{({\epsilon^{c}}_{\perp}+\epsilon_{m}) ({\epsilon^{c}}_{\parallel}-\epsilon_{m})}
\label{eq:adef}
\end{equation}
and is obviously frequency dependent. Parameters $a_1$ and $a_2$ can be thought of as a specific measure of the anisotropy of the cylinders in the left and right half-spaces when 
compared with the isotropic bathing medium $m$. Note that they vanish when the transverse dielectric response of the cylinder material equals the medium response. 
The explicit form of the second derivative of $f(\ell,\theta)$ now follows as 

\begin{widetext}
\begin{eqnarray}
\frac{d^{2}f(\ell,\theta)}{d\ell^{2}} &=& - \frac{v_1 v_2 \Delta_{1,\parallel} \Delta_{2,\parallel}}{32} 
\frac{e^{-2 \ell \sqrt{Q^{2} + \epsilon_m \frac{\omega_n^{2}}{c^{2}}}}}{(Q^{2} + \epsilon_m \frac{\omega_n^{2}}{c^{2}})} \nonumber \\
& &
\left\{ 2 \left[ (1+3a_1)(1+3a_2) Q^{4} + 2 (1+2a_1+2a_2+3a_1a_2) Q^{2} \epsilon_m \frac{\omega_n^{2}}{c^{2}} + 2(1+a_1)(1+a_2) {\epsilon_m}^{2} \frac{\omega_n^{4}}{c^{4}}\right] \right. + \nonumber\\
& & \left. ~~~~~~~~~~~~~~~~~~~~~~~~~~~~~~~~~~~~~~~~~~~~~~~~~~~~ + (1-a_1)(1-a_2)\left( Q^{2} + 2 \epsilon_m \frac{\omega_n^{2}}{c^{2}} \right)^2 \cos 2\theta \right \}.
\label{eq:d2f}
\end{eqnarray}
\end{widetext}
Here $R_1$ and $R_2$ are the cylinder radii, assumed to be the smallest lengths in the problem \cite{Barash89}. The frequency dependence of the dielectric functions is in 
$\epsilon_m(i \omega_n)$, ${\epsilon^{c}}_{\perp}(i \omega_n)$ and ${\epsilon^{c}}_{\parallel}(i \omega_n)$, and therefore also $a = a(i \omega_n)$. The frequencies 
in the Matsubara summation are $\omega_n = 2\pi~\frac{k_BT}{\hbar} n$. Note that Eq. \ref{eq:d2f} is symmetric with respect to 1 and 2 indices (left and right 
half-spaces), as it should be.

This is as far as a general theory can go. We must now deal separately with the cases of skewed and parallel cylinders, since the connection 
between $\frac{d^{2}{\cal G}(\ell,\theta)}{d\ell^{2}}$ and the effective pair interaction between cylinders is different for the two cases, 
see Eqs. \ref{form1},\ref{form2}. We first analyze the case of skewed cylinders.

\subsection{Skewed cylinders}

We use  Eq. \ref{form1} to obtain the interaction free energy between two skewed cylinders:
\begin{equation}
G(\ell,\theta) = - \frac{k_BT}{64 \pi} \frac{ \pi^2 R_1^{2} R_2^{2} }{\ell^{4} \sin{\theta}} {\sum_{n=0}^{\infty}}' \Delta_{1,\parallel} \Delta_{2,\parallel} \int_0^{\infty}  u du ~\frac{e^{- 2 \sqrt{u^{2} + p_n^{2}}}}{(u^{2} + p_n^{2})}  g(a_1, a_2, u, p_n, \theta),
\label{pars-3}
\end{equation}
where $u = Q \ell$,  
\begin{widetext}
\begin{eqnarray}
g(a_1, a_2, u, p_n, \theta) &=&  2 \left[ (1+3a_1)(1+3a_2) u^{4} + 2(1+2a_1+2a_2+3a_1a_2) u^{2}p_n^{2}  + 2(1+a_1)(1+a_2) p_n^{4}\right] + \nonumber \\ 
& & ~~~~~~~~~~~~~~~~~~~~~~~~~~~~~~~~~~~~~~~~~ + (1-a_1)(1-a_2)(u^{2} + 2 p_n^{2})^2 \cos 2\theta
\end{eqnarray}
\end{widetext}
and $p_n^{2} =  \epsilon_m(i \omega_n) \frac{\omega_n^{2}}{c^{2}} \ell^{2}$. Another change of variables with $u = p_n t$, yields 
\begin{equation}
G(\ell,\theta) = - \frac{k_BT}{64 \pi} \frac{ \pi^2 R_1^{2} R_2^{2} }{\ell^{4} \sin{\theta}} {\sum_{n=0}^{\infty}}' \Delta_{1,\parallel} \Delta_{2,\parallel} ~p_n^{4} ~\int_0^{\infty} t dt ~\frac{e^{- 2 p_n \sqrt{t^{2} + 1}}}{(t^{2} + 1)} \tilde g(t, a_1(i \omega_n), a_2(i \omega_n), \theta),
\label{pars-31}
\end{equation}
with
\begin{widetext}
\begin{eqnarray}
\tilde g(t, a_1, a_2, \theta) &=& 2 \left[ (1+3a_1)(1+3a_2) t^{4} + 2 (1+2a_1+2a_2+3a_1a_2) t^{2}  + 2(1+a_1)(1+a_2)\right] + \nonumber \\
& & ~~~~~~~~~~~~~~~~~~~~~~~~~~~~~~~~~~~~~~~~~ + (1-a_1)(1-a_2)(t^{2} + 2)^2 \cos 2\theta.
\end{eqnarray}
\end{widetext}
This is the final result for the cylinder-cylinder interaction at all angles when the radii of the cylinders are the smallest lengths in the system. It includes retardation and the full angular dependence. Some simple limits that can be obtained form this general expression.

The  non-retarded limit where $c \longrightarrow \infty$, has already been explored in Ref. \onlinecite{rick2}. There $p_n \longrightarrow 0$ for all $n$ and we obtain from Eq. \ref{pars-3}
\begin{eqnarray}
G(\ell,\theta; c \longrightarrow \infty) &=& - \frac{k_BT}{64 \pi} \frac{\pi^{2} R_1^{2} R_2^{2} }{\ell^{4} \sin{\theta}} {\sum_{n=0}^{\infty}}' \Delta_{1,\parallel} \Delta_{2,\parallel}
\int_0^{\infty} u^{3} du ~{e^{- 2 u}} \left[ 2 (1+3a_1)(1+3a_2) + (1-a_1)(1-a_2)  \cos 2\theta \right] = \nonumber\\
&=& - \frac{k_BT}{64 \pi} \frac{\pi^{2} R_1^{2} R_2^{2}}{\ell^{4} \sin{\theta}} {\sum_{n=0}^{\infty}}' \Delta_{1,\parallel} \Delta_{2,\parallel}
 \frac{3}{8}\left[ 2 (1+3a_1) (1+3a_2) + (1-a_1) (1-a_2)  \cos 2\theta \right].
\label{pars-4}
\end{eqnarray}
This formula could also be obtained directly from Eq. \ref{pars-31} taking into account that in the $t$ integration only the terms with large $t$ contribute to the final integral. 
Expanding the whole integrand for large $t$ returns us to Eq. \ref{pars-4}. 

At low temperatures, when the summation over the Matsubara frequencies can be turned into an integral over $n$ with $dn = \hbar/(2\pi ~k_BT) d\omega$, the corresponding interaction free energy is  
\begin{equation}
G(\ell,\theta) = - \frac{\hbar}{128 \pi^2} \frac{\pi^2 R_1^{2} R_2^{2}}{c^{4} \sin{\theta}} \int_{0}^{\infty}\!\!\! d\omega~ \omega^{4}  
\Delta_{1,\parallel}(i\omega) \Delta_{2,\parallel}(i\omega) \epsilon_m(i\omega)^2\!\!\!\int_0^{\infty}\!\!\! t dt ~\frac{e^{- 2 \sqrt{\epsilon_m(i \omega)} 
\frac{\omega}{c} \ell \sqrt{t^{2} + 1}}}{(t^{2} + 1)} \tilde g(t, a_1(i \omega), a_1(i \omega), \theta).
\label{pars-32}
\end{equation}
We now rework this equation to obtain the retarded result for the interaction between two semiconducting cylinders. Note here that we can not derive the Casimir limit properly  
as our formulation is not valid for nominally infinite zero-frequency (Drude-like) dielectric response. For that case see  Ref. \onlinecite{Barash89}. First instead of  variable $\omega$, we introduce 
$x = \frac{\ell}{c} \sqrt{t^2 + 1} ~\omega$. Then, following closely the arguments in Ref. \onlinecite{LL} we obtain  the interaction free energy in the form
\begin{equation}
G(\ell,\theta) = - \frac{\hbar c}{128 \pi^2} \frac{\pi^2 R_1^{2} R_2^{2}} { \ell^5\sin{\theta}} \epsilon_m(0)^2 \Delta_{1,\parallel}(0) \Delta_{2,\parallel}(0) 
\int_{0}^{\infty}\!\!\! dx~ x^{4}   \!\!\!\int_0^{\infty}\!\!\! t dt ~\frac{e^{- 2 \sqrt{\epsilon_m(0)}x}}{(t^{2} + 1)^{7/2}}~ \tilde g(t, a_1(0), a_2(0), \theta).
\label{pars-33}
\end{equation}
Here $\epsilon_m(0)$ and $a_1(0), a_2(0)$ denote the static, {\sl i.e.} zero frequency, values of the corresponding functions. 
Obviously in this regime the interaction free energy decays faster with separation, being a reflection of the retardation. All the frequency 
dependence of the material properties is reduced to the static response in this limit, just as in the general Lifshitz analysis \cite{LL}.

\subsection{Parallel cylinders}

The analysis here is somewhat more complicated because the pair interaction energy between the cylinders involves the inverse Abel transform. We start with
\begin{equation}
\frac{d^{2}{\cal G}(\ell,\theta = 0)}{d\ell^{2}} = \frac{k_BT}{2\pi} {\sum_{n=0}^{\infty}}' \int_0^{\infty} Q dQ \frac{d^{2}f(\ell,\theta = 0)}{d\ell^{2}},
\label{equnew1}
\end{equation}
where  
\begin{eqnarray}
\frac{d^{2}f(\ell,\theta = 0)}{d\ell^{2}} &=& - \frac{v_1 v_2 \Delta_{1,\parallel} \Delta_{2,\parallel}}{32} 
\frac{e^{-2 \ell \sqrt{Q^{2} + \epsilon_m \frac{\omega_n^{2}}{c^{2}}}}}{(Q^{2} + \epsilon_m \frac{\omega_n^{2}}{c^{2}})} \nonumber \\
& & \left \{ 2 \left [ (1+3a_1)(1+3a_2) Q^{4} + 2 (1+2a_1+2a_2+3a_1a_2) Q^{2} \epsilon_m \frac{\omega_n^{2}}{c^{2}} + 2(1+a_1)(1+a_2) 
{\epsilon_m}^{2} \frac{\omega_n^{4}}{c^{4}}\right] \right. + \nonumber\\
& &~~~~~~~~~~~~~~~~~~~~~~~~~~~~~~~~~~~~~~~~~~~~~ +  \left. (1-a_1)(1-a_2)(Q^{2} + 2 \epsilon_m \frac{\omega_n^{2}}{c^{2}})^2 \right \},
\label{equnew2}
\end{eqnarray}
and again $v_1 = N~\pi R_1^{2}$ ($v_2 = N~\pi R_2^{2}$) and $a = \frac{2 \Delta_{\perp}}{\Delta_{\parallel}}$. We continue by introducing the Abel transform and its properties. Namely, if we define
\begin{equation}
\int_{-\infty}^{+\infty}g(\sqrt{\ell^{2}+y^{2}})~dy = f(y),
\end{equation} 
then
\begin{equation}
g(\ell) = - \frac{1}{\pi} \int_{\ell}^{+\infty} \frac{f'(y) dy}{\sqrt{y^2 - \ell^2}}.
\end{equation} 
Taking this into account when considering Eqs. \ref{equnew2}, we remain with 
\begin{equation}
g(\ell) = - \frac{k_BT}{32} {R_1^{2} R_2^{2}} 
{\sum_{n=0}^{\infty}}' \Delta_{1,\parallel} \Delta_{2,\parallel} \int_{\ell}^{+\infty}\!\!\!\!\! \frac{dy}{\sqrt{y^2 - \ell^2}} \int_0^{\infty}\!\!\!  
Q dQ \frac{e^{-2 y \sqrt{Q^{2} + \epsilon_m(i \omega_n) \frac{\omega_n^{2}}{c^{2}}}}}{(Q^{2} + \epsilon_m(i \omega_n) \frac{\omega_n^{2}}{c^{2}})^{1/2}} 
h(a_1(i \omega_n), a_2(i \omega_n), Q, \epsilon_m(i \omega_n) \frac{\omega_n^{2}}{c^{2}}),
\end{equation} 
where
\begin{eqnarray}
h(a_1, a_2, Q, \epsilon_m \frac{\omega_n^{2}}{c^{2}}) &=&  2 \left[ (1+3a_1)(1+3a_2) Q^{4} + 2 (1+2a_1+2a_2+3a_1a_2) Q^{2} 
\epsilon_m \frac{\omega_n^{2}}{c^{2}} + 2(1+a_1)(1+a_2) {\epsilon_m}^{2} \frac{\omega_n^{4}}{c^{4}}\right] +  \nonumber \\ 
& & (1-a_1)(1-a_2)(Q^{2} + 2 \epsilon_m \frac{\omega_n^{2}}{c^{2}})^2 .
\end{eqnarray}
As before, we introduce $p_n^{2} =  \epsilon_m(i \omega_n) \frac{\omega_n^{2}}{c^{2}} \ell^{2}$, $u = Q\ell$ and $y \longrightarrow y/\ell$. This allows us to rewrite the above integrals as
\begin{equation}
g(\ell) = - \frac{k_BT}{32} \frac{R_1^{2} R_2^{2}}{\ell^5} {\sum_{n=0}^{\infty}}' \Delta_{1,\parallel} \Delta_{2,\parallel} 
\int_{1}^{+\infty}\!\!\!\!\! \frac{dy}{\sqrt{y^2 - 1}} \int_0^{\infty}\!\!\!  u du ~\frac{e^{-2 y \sqrt{u^{2} + p_n^{2}}}}{(u^{2} +p_n^{2})^{1/2}} ~h(a_1(i \omega_n), a_2(i \omega_n), u, p_n^{2}),
\label{retardedfinal}
\end{equation}
and
\begin{eqnarray}
h(a_1(i \omega_n), a_2(i \omega_n), u, p_n^{2}) &=&   2 \left[ (1+3a_1)(1+3a_2) u^{4} + 2 (1+2a_1+2a_2+3a_1a_2) u^{2} p_n^{2} + 2(1+a_1)(1+a_2) p_n^{4}\right]  + \nonumber \\
& &  (1-a_1) (1-a_2) (u^{2} + 2 p_n^{2})^2 .
\end{eqnarray}
This is the final result for the interaction between two parallel thin cylinders at all separations and contains retardation effects explicitly.  In general, the above expression can only be evaluated numerically once the dielectric spectra of component substances are known.

In the non-retarded limit, $c \longrightarrow \infty$, the above formula  reduces to 
\begin{eqnarray}
g(\ell; c \longrightarrow \infty) &=& - \frac{k_BT}{32} \frac{ R_1^{2} R_2^{2} }{\ell^5} {\sum_{n=0}^{\infty}}' \Delta_{1,\parallel} \Delta_{2,\parallel} 
\left( 3 + 5 (a_1+a_2) + 19 a_1 a_2 \right) \int_{1}^{+\infty}\!\!\!\!\! \frac{dy}{\sqrt{y^2 - 1}} \int_0^{\infty}\!\!\! u^4 du ~{e^{-2 y u }} = \nonumber\\
& & - \frac{9 ~k_BT}{(64 \times 32) \pi } \frac{\pi^2 R_1^{2} R_2^{2}}{\ell^5} {\sum_{n=0}^{\infty}}' \Delta_{1,\parallel} \Delta_{2,\parallel}
\left \{ 3 + 5 [ a_1(i \omega_n) + a_2(i \omega_n) ] + 19 a_1(i \omega_n) a_2(i \omega_n) \right \}.
\label{eq:cyl-paral-nonretarded}
\end{eqnarray}
For the case where the two interacting cylinders are composed of solid isotropic dielectric materials this form of the interaction free energy can be compared with the result obtained by Barash and Kyasov ( Eq. 10 in \cite{Barash89}) and can be reduced to it {\sl exactly}.

As with skewed cylinders, we can take the zero temperature limit where the summation over the Matsubara frequencies becomes an integral over $n$ with $dn = \hbar/(2\pi ~k_BT) d\omega$. Again we introduce $x = \frac{\ell}{c} \sqrt{t^2 + 1} ~\omega$. Then, as for skewed cylinders, we obtain the interaction free energy per unit length of two parallel cylinders,
\begin{equation}
g(\ell) = -\frac{\hbar c}{64 \pi^3} \frac{\pi^2 R_1^{2} R_2^{2}}{\ell^6}  \epsilon_m(0)^{5/2} \Delta_{1,\parallel}(0) \Delta_{2,\parallel}(0) 
\!\! \int_0^{\infty} dx~ x^5 \int_{1}^{+\infty}\!\!\!\!\! \frac{dy}{\sqrt{y^2 - 1}} \int_0^{\infty} \frac{t dt ~e^{- 2 \sqrt{\epsilon_m(0)} ~y x}}{(t^2 + 1)^{7/2}} \tilde h(t, a_1(0), a_2(0)).
\end{equation}
Here 
\begin{equation}
\tilde h(t, a_1, a_2) =   2 \left[ (1+3a_1)(1+3a_2) t^{4} + 2 (1+2a_1+2a_2+3a_1a_2) t^{2}  + 2(1+a_1)(1+a_2)\right] +  (1-a_1)(1-a_2)(t^{2} + 2)^2 .
\end{equation}
The spatial dependence is, again, one power higher in the retarded regime than in the non-retarded regime. All the frequency dependence of the material properties in the retarded limit is again reduced to the static response as in the Lifshitz analysis \cite{LL}.

\section{Numerical results: retarded vs. nonretarded interaction}

Expressions for the van der Waals - dispersion interaction free energy between parallel and skewed cylinders derived above can be analysed numerically, once the spectral properties of the interacting cylinders are given. In all the results in this section, the medium between cylinders is assumed to be vacuum, $\epsilon_3 (i \omega) = 1$, and the temperature is $T=297$ K. 
The spectral properties of the cylinders are taken to be those of thin single walled nanotubes as calculated by {\sl ab initio} methods in the optical range \cite{rick2}.
Robust quantum mechanical codes have been developed to give us the very accurate spectral data needed to investigate the numerical consequences of the above theory. Without going into details, we should note that the dielectric spectral data depend on chirality and some SWCNTs even exhibit significant optical anisotropy between their radial and axial directions \cite{Mintmire_CNT_overview}. For purposes of illustration of the theory developed here, we choose [5,1] and [29,0] semiconducting carbon nanotubes, which have been previously analyzed and which differ substantially in their radii as well as spectral properties \cite{Rajter_e2_LD}. This choice is motivated by the fact that although [4,2] is technically the smallest SWCNT - it has a radius of only 0.207 nm - the [5,1] and [29,0] have a bigger difference in their spectra and are thus more appropriate to explore the effect of disparate dielectric spectra on van der Waals - dispersion forces.

The theory developed here should be safely applied to this case since the tubes are not metallic and their dielectric response functions are thus always bounded. Furthermore, their dielectric responses and radii (1.135 and 0.218 nm for [29,0] and [5,1] SWCNTs, respectively, when measured from the center of the cylinder to the centers of the carbon atoms) are very different, which is another reason for choosing these tubes as the benchmark for application of the theory. For illustrative purposes we also disregard the finite core size of the [29,0] SWCNT that would require a more careful modeling of its effective dielectric response \cite{rick1} and thus introduce additional parameters that would complicate the understanding of the retardation effects in van der Waals - dispersion interactions between these two SWCNTs, which is our primary aim in this paper. Once the surface to surface separation between two SWCNTs is greater than approximately two SWCNT outer diameters \cite{rick1} this approximation turns out to work quite well.

\subsection{Parallel cylinders}
\begin{figure}[ht]
\centerline{
\epsfig {file=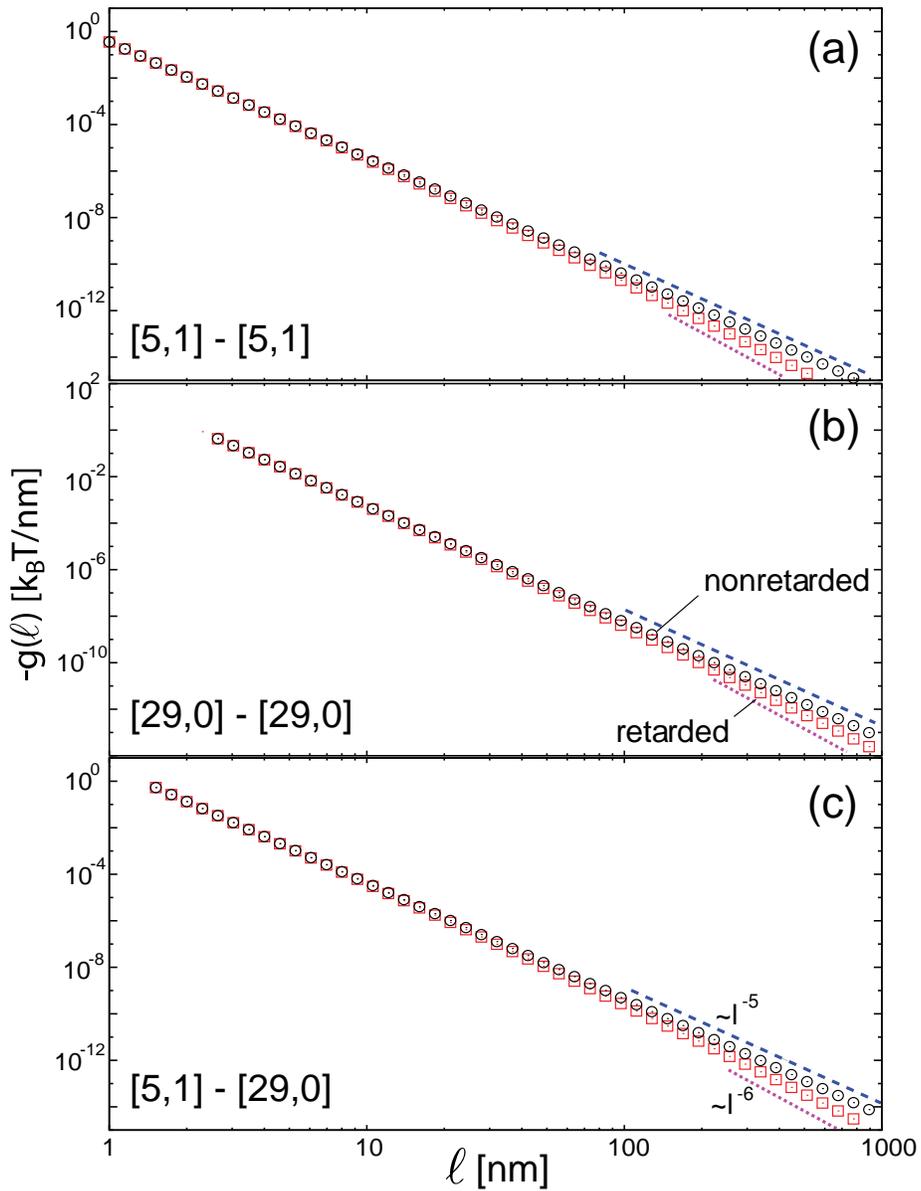,width=12cm}
}
\caption{The van der Waals interaction free energy per unit length between two parallel [5,1] SWCNTs (panel a), two parallel [29,0] SWCNTs (panel b), and 
parallel [5,1] and [29,0] SWCNTs (panel c) as a function of their separation. Circles represent the nonretarded expression \cite{rick1}, while squares represent 
the fully retarded calculation Eq. \ref{retardedfinal}. Dashed and dotted lines indicate $\ell^{-5}$ and $\ell^{-6}$ dependences, respectively. The interaction free energy is calculated only for interaxial separations larger then $\ell = R_1 + R_2$, {\sl i.e.} in the non-intersecting regime of the cylinders. Close to this value the thin cylinder approximation breaks down and is superseded by a different type of calculation, see \cite{rick2} for details.
}
\label{fig:fig1}
\end{figure}

It seems plausible that retardation effects are largest in the parallel configuration. 
Although the calculations are most demanding in this case, we nevertheless analyze it first. 
Figure \ref{fig:fig1} shows the interaction free energy per unit length for two parallel [5,1] SWCNTs (panel a), two parallel [29,0] SWCNTs (panel b) and 
parallel [5,1] and [29,0] SWCNTs (panel c).

It is obvious, Fig. \ref{fig:fig1}, that the departure of the retarded results from the 
nonretarded ones takes place only when separations are larger than about 50 nm, but the 
strength of van der Waals interaction there is weak, on the order of 10$^{-8}$ $k_B T$ / nm \footnote{Note the equivalence $1k_B T \sim 4zJ$.} for 
[5,1]-[29,0] pair of SWCNTs. One can see very gradual deviation of the retarded results from $\ell^{-5}$ to $\ell^{-6}$ scaling with the separation distance (note log-scale on both axes). 

For $\ell=2$ nm, the fully retarded value of the van der Waals interaction for 
the [5,1]-[29,0] pair is -0.13294 $k_B T$ / nm, while the nonretarded value is -0.13377 $k_B T$ / nm, so 
the contribution of retardation at this distance is only about 0.6 \%. The nonretarded value of 
interaction then is of some use for a large interval of intercylinder separations due to the strict power-law behavior of the nonretarded values $\propto \ell^{-5}$. 
For example, for $\ell=16$ nm, the interaction is -0.13377 (2/16)$^5$ = -4.0823 10$^{-6}$ $k_B T$ / nm. The full retarded expressions give a mostly negligible contribution to the van der Waals interactions at spacings of relevance. Its importance can be boosted when considering interactions between different types of SWCNTs in a dielectric medium with appropriate (or even ''tailored'') response. In that case and depending on the full dielectric spectra of the two cylinders and of the medium, retardation coupled to the dielectric spectra can in principle lead to a change in sign of the interaction via a similar mechanism first considered for the interaction of ice with vacuum across a liquid film, as studied by Elbaum and Schick \cite{Elbaum}. (In that case, the two half-spaces are ice and vacuum, while the dielectric medium between is water). These effects will be investigated in Sec. \ref{sec:tailor}. Note that the sign change can also be obtained for hollow cylinders at small separations, but it has a different origin \cite{Rajter_e2_LD}.

Of all the three cases considered, the van der Waals attraction is the strongest for the pair of [29,0] SWCNTs (-0.5325 $k_B T$/nm at $\ell=4$ nm). The most important reason for this is the explicit dependence of the van der Waals interaction on radii of the two cylinders ($R_1^{2} R_2^{2}$). 
The radius of [29,0] SWCNT is 5.21 times bigger than the radius of [5,1] SWCNT.

\subsection{Skewed cylinders}

\begin{figure}[ht]
\centerline{
\epsfig {file=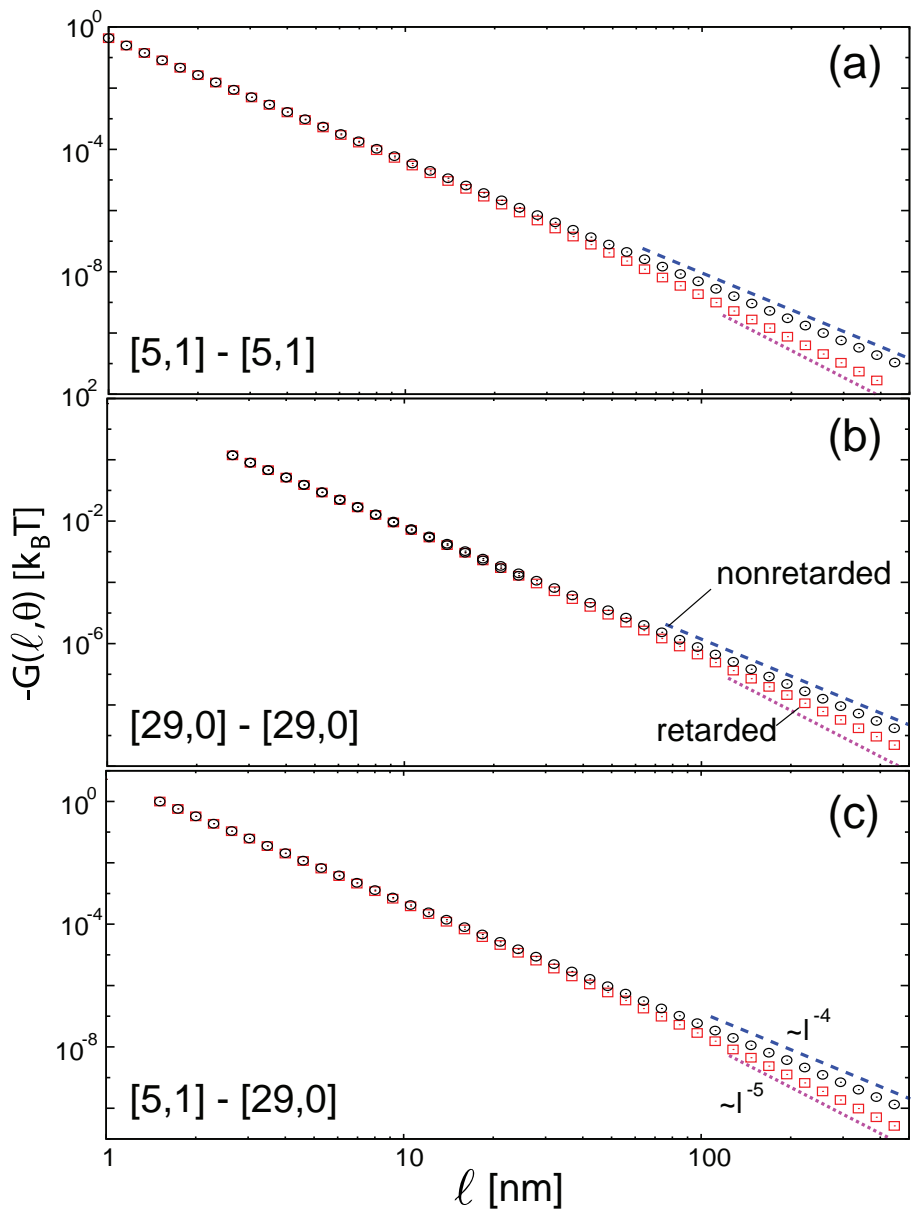,width=12cm}
}
\caption{The van der Waals interaction free energy for two [5,1] SWCNTs (panel a), two [29,0] SWCNTs (panel b) and [5,1] and [29,0] SWCNTs (panel c) crossed at the right angle ($\theta = \pi / 2$) as a function of their separation. Circles represent the nonretarded expression \cite{rick1}, while squares represent the fully retarded calculation Eq. \ref{retardedfinal}. Dashed and dotted lines indicate $\ell^{-4}$ and $\ell^{-5}$ dependences, respectively. The interaction free energy is calculated only for interaxial separations larger then $\ell = R_1 + R_2$, {\sl i.e.} in the non-intersecting regime of the cylinders. Close to this value the thin cylinder approximation breaks down and is superseded by a different type of calculation, see \cite{rick2} for details.
}
\label{fig:fig2}
\end{figure}

These calculations are numerically less demanding since the dimensionality of 
integration is one rather then two for  parallel cylinders. Figure \ref{fig:fig2} 
presents the dependence of the van der Waals interaction free energy for two [5,1] SWCNTs (panel a), two [29,0] SWCNTs (panel b) and [5,1] and [29,0] SWCNTs (panel c) crossed at the right angle ($\theta = \pi / 2$). 
Again, the contribution of retardation is small. Its observable effects in this 
configuration take place at about 30 nm, similar as in the case of parallel nanotubes. The nonretarded 
and the retarded values for the van der Waals interaction for [5,1]-[29,0] pair separated by $\ell=2$ nm 
are -0.3255 $k_B T$ and -0.3285 $k_B T$, respectively. The contribution of retardation effects to the van der Waals interaction at this distance is thus 0.9 \%, similar to the case of parallel SWCNTs, and, interestingly, somewhat larger. The dependence of the pair interaction on separation is $\propto \ell^{-4}$ for the nonretarded case, but one can see a gradual transition towards the $\propto \ell^{-5}$ dependence for large distances. Again, the interaction is strongest for a pair of [29,0] carbon nanotubes.

\begin{figure}[ht]
\centerline{
\epsfig {file=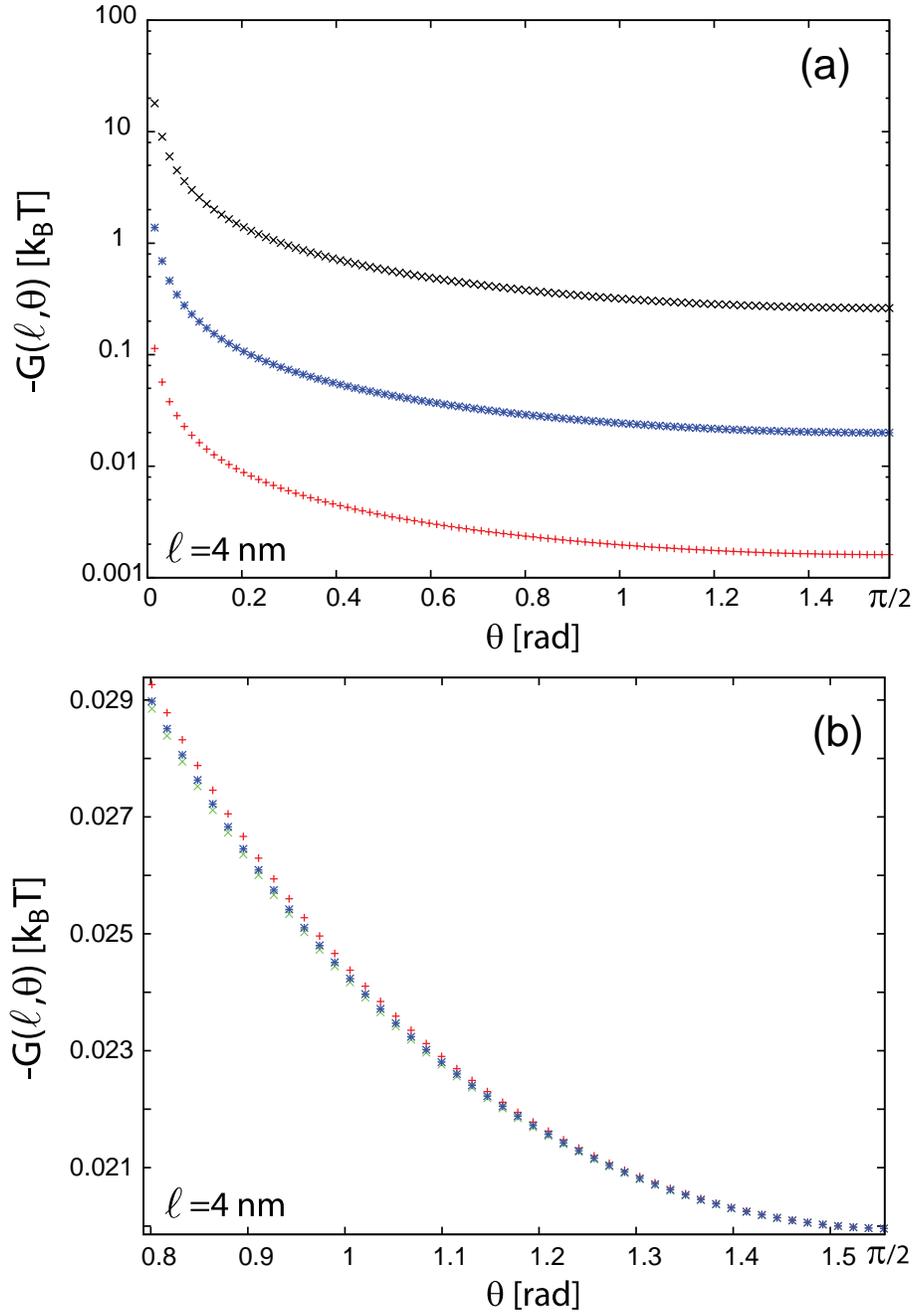,width=12cm}
}
\caption{Panel a): The van der Waals interaction energy between two two [5,1] SWCNTs (pluses), two [29,0] SWCNTs (x-es) and 
[5,1] and [29,0] SWCNTs (stars) separated by $\ell=4$ nm as a function of their relative angle (fully retarded calculation).
Panel b): The same as in panel a), only the energies have been rescaled so that the three cases yield the same value for 
$\theta = \pi/2$.
}
\label{fig:fig3}
\end{figure}

The dependence of the van der Waals interaction on the relative orientation angle ($\theta$) is of interest, since it can be used to calculate the effective van der Waals torque that acts to make the 
cylinders parallel. The dependence of the van der Waals interaction on the angle for $\ell=4$ nm is 
shown in Fig. \ref{fig:fig3}
Note that the torque vanishes for $\theta=\pi / 2$. One should also note that for $\theta = 0$ the van der Waals interaction 
calculated from Eq. \ref{pars-31} diverges. This is due to the fact the the interaction free energy for parallel cylinders scales 
with their length and thus diverges for infinitely long cylinders. It is clear from Eqs. \ref{pars-31} and \ref{pars-4} that the 
angular dependence of the van der Waals interaction depends on the details of the dielectric response of the cylinders. Nevertheless, 
a quick look at panel a) of Fig. \ref{fig:fig3} may lead one to think that the angular dependencies for all three cases are 
completely the same up to some multiplicative constant. This is in part due to the divergence introduced by the $1/\sin \theta$ factor which 
tends to screen the fine details of the interaction. However, a closer look indeed shows that the angular anisotropies are not 
scalable as is shown in panel b) of Fig. \ref{fig:fig3}, where all three interactions have been scaled so as to reproduce the 
value of [5,1] - [29,0] interaction at $\theta = \pi /2$ (this means that the scaling factor for [5,1] - [29,0] interaction 
is 1). One can see how the curves separate as $\theta$ diminishes, and that 
[5,1] - [5,1] case has the largest angular anisotropy of the van der Waals interaction. The effect is relatively small, however, and the $1/\sin \theta$ factor determines the anisotropy to the largest extent, at least for the cases considered here.

%This divergence is not physical 
%and is due to the fact that finite size effects are not properly accounted fro in a calculation based on the Pitaevskii {\sl ansatz}.
%I DON'T GET THIS RUDI... I THINK THAT EVERYTHING IS OK.

\section{The influence of dielectric response of bathing medium on the van der Waals interaction: repulsive, attractive, or both?}
\label{sec:tailor}

An interesting question is whether the cylinder-cylinder interaction can be repulsive and how that depends on the dielectric 
responses of the cylinders and the bathing medium (this can be achieved only when the cylinders are different). 
The answer is somewhat hidden in Eq. \ref{retardedfinal} and much more 
explicit in Eq. \ref{eq:cyl-paral-nonretarded}. In both of these equations, the van der Waals interaction depends on the product 
$\Delta_{1,\parallel} \Delta_{2,\parallel}$ which appears under the sum over Matsubara frequencies. There is, however, the 
additional term that multiplies it [$3+5(a_1+a_2)+19a_1a_2$ for parallel cylinders in the nonretarded regime], that depends on 
both the longitudinal and transverse polarizabilities 
of the cylinders. A closer examination of Eq. \ref{eq:adef} shows that for isotropic cylinders $a$ is neccessarily positive, 
so that the aditional term is also positive when both cylinders are isotropic, at least in the nonretarded regime. 
Parameter $a$ is negative only when the dielectric response of the medium is between the longitudinal and transverse 
responses of the cylinder, and even then, 
the sign of the additional term depends on the details of response of other cylinder. It thus seems that the repulsive nature 
of van der Waals interaction can be most easily obtained and comprehended when it is enforced through the longitudinal response of the 
two cylinders. Assuming now that the 
terms that multiply $\Delta_{1,\parallel} \Delta_{2,\parallel}$ (two-dimensional integral in the retarded case and relatively 
simple combination of transverse and logitudinal dielectric responses in the nonretarded case) are positive, one concludes 
that the repulsive van der Waals interaction can be obtained by making the product $\Delta_{1,\parallel} \Delta_{2,\parallel}$ 
negative for all imaginary frequencies, i.e. that (see Eq. \ref{anisoind})
\begin{equation}
[\epsilon^c_{1,\parallel}(i \omega)-\epsilon_m(i \omega)] [\epsilon^c_{2,\parallel}(i \omega)-\epsilon_m (i \omega)] < 0, \; \forall \omega
\end{equation}
(a completely analogous finding has been recently verified experimentally for the repulsive interaction between gold sphere
and silica plate immersed in bromobenzene \cite{ParsegianNature}). 
This leads us to an interesting conception of ''designing'' the van der Waals interaction between the cylinders by the introduction of an 
appropriate intervening medium. A simplest way which should produce repulsive interaction 
(at least in the $\Delta_{1,\parallel} \Delta_{2,\parallel}$ ''channel'') is to ''construct'' the medium dielectric response as
\begin{equation}
\epsilon_m(i \omega) = \frac{\epsilon^c_{1,\parallel}(i \omega) + \epsilon^c_{2,\parallel}(i \omega)}{2}.
\label{mixedresponse}
\end{equation}
Since dielectric responses are difficult to measure experimentally the "designer" response above would have to be engineered by the use of {\sl ab initio} codes to fill this gap.

\begin{figure}[ht]
\centerline{
\epsfig {file=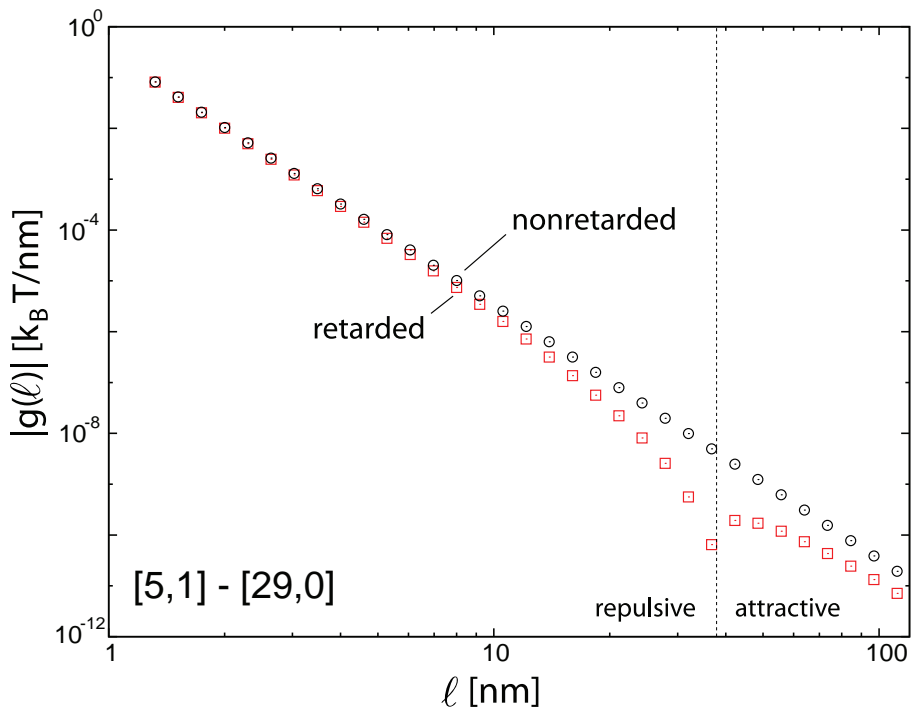,width=12cm}}
\caption{The absolute value of the retarded (squares) and nonretarded (circles) van der Waals 
interaction energy between parallel [5,1] and [29,0] SWCNTs as a function of their separation $\ell$. The intervening medium is described as Eq. \ref{mixedresponse}. The retarded van der Waals interaction is repulsive when $\ell < 35$ nm and attractive when $\ell > 35$ nm.}
\label{fig:fig4}
\end{figure}

The assumption that the above dielectric response of the intervening medium leads to repulsive dispersion interactions between two SWCNTs can be easily checked numerically and, as shown in Fig. \ref{fig:fig4}, the van der Waals interaction between [5,1] and [29,0] SWCNTs is indeed repulsive, at small distances $\ell$, but an interesting effect takes place when $\ell \approx 35$ nm. Namely the retarded van der Waals interaction changes sign, becoming attractive when $\ell > 35$ nm and with an 
extremely shallow minimum ($\sim 10^{-10} k_B T$/nm) at $\ell \sim$ 36 nm. This effect is not seen in the nonretarded 
van der Waals interaction which is repulsive for all distances $\ell$.
Although we aimed at constructing the medium response that shall guarantee the repulsive van der Waals interaction, we obtained that 
the retardation effects can act so as to change the sign of interaction at a certain separation distance. This effect is quite 
similar to what has been found in Ref. \onlinecite{Elbaum} and it is of interest to see whether it can be ''boosted'' and 
brought to smaller separation distances by a carefully guided choice of the medium in between the cylinders.

Retardation acts so as to screen the contribution of higher Matsubara frequencies to the total value of the van der Waals interaction. 
This effect becomes more important as the separation distances increase \cite{Parsegian}. It thus seems possible to design a dielectric 
response of the medium so that the retardation screens the repulsive large $n$ contributions to the 
summation in Eq. \ref{retardedfinal}, switching from repulsive to attractive behavior for some $\ell$. One can even imagine 
a medium in which the van der Waals interaction between the cylinders would be attractive for small $\ell$'s and repulsive for large 
$\ell$'s. This shall be illustrated by several examples.

\begin{figure}[ht]
\centerline{
\epsfig {file=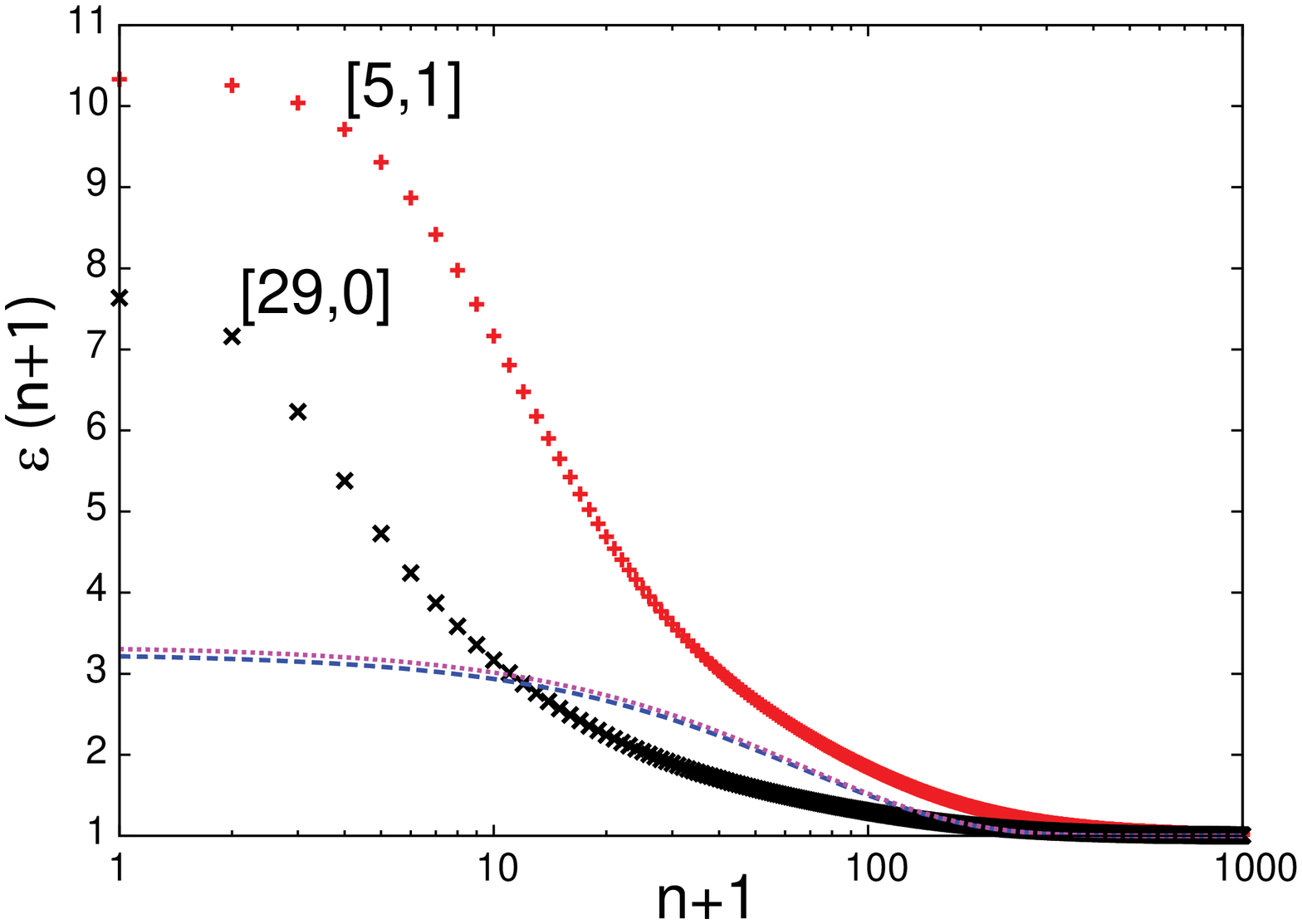,width=12cm}
}
\caption{Longitudinal dielectric responses of [5,1] (pluses) and [29,0] (x-es) SWCNTs as a function of the Matsubara frequency index. 
Two model medium responses are indicated by dashed and dotted lines (see text).}
\label{fig:figodzivi}
\end{figure}

\begin{figure}[ht]
\centerline{
\epsfig {file=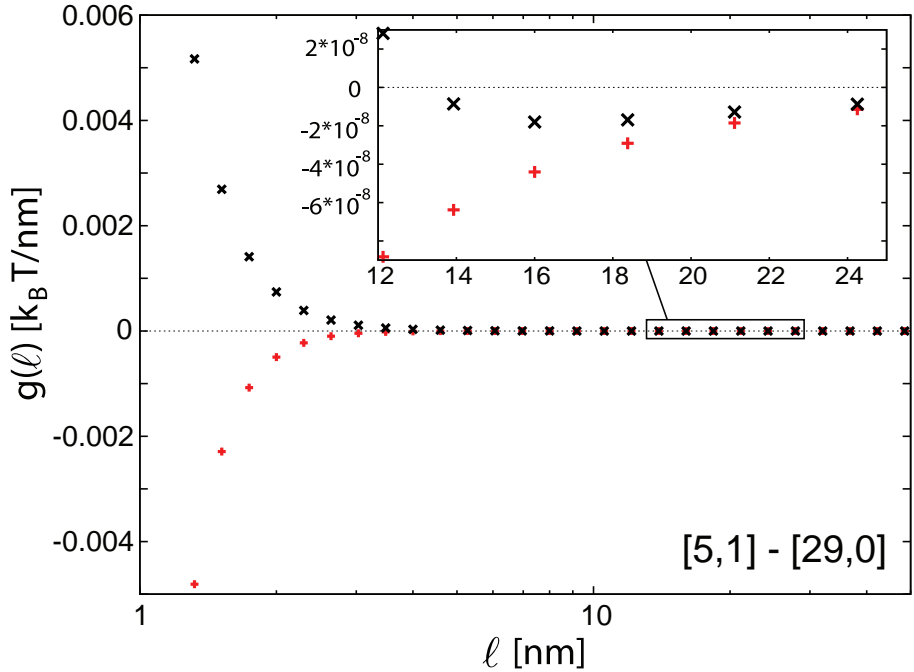,width=12cm}
}
\caption{Retarded van der Waals interaction energy per unit length between parallel [5,1] and [29,0] SWCNTs. Pluses (x-es) indicate the 
results for the medium response shown by dashed (dotted) line in Fig. \ref{fig:figodzivi}.}
\label{fig:fig5}
\end{figure}

\begin{figure}[ht]
\centerline{
\epsfig {file=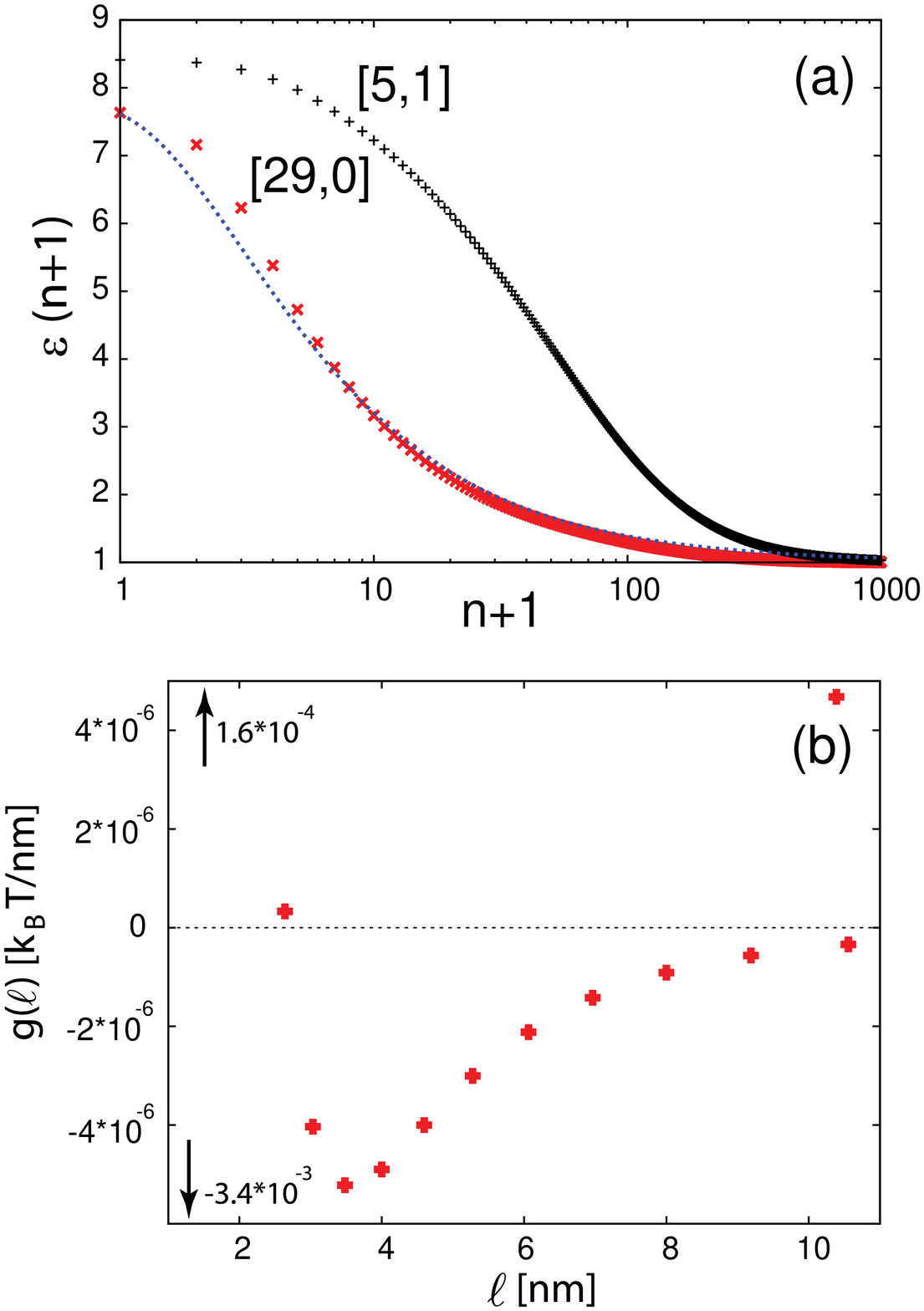,width=12cm}
}
\caption{Panel (a) represents the longitudinal dielectric responses of [5,1] SWCNTs (pluses), [29,0] SWCNTs (x-es), and a 
hypothetical medium (dashed line) as a function of the Matsubara frequency index. Panel (b) shows the (retarded) 
van der Waals interaction energy per unit length between parallel [5,1] and [29,0] SWCNTs in such a medium.}
\label{fig:figmednov}
\end{figure}

Figure \ref{fig:figodzivi} shows the longitudinal dielectric responses of [5,1] and [29,0] SWCNTs. The two model medium responses 
are indicated by dashed and dotted lines, respectively. The two responses are modeled as $\epsilon_m (n) = 1 + 2.215 \exp(-0.015n)$ 
(dashed line, model 1) and $\epsilon_m (n) = 1 + 2.305 \exp(-0.015n)$ (dotted line, model 2), where $n$ is the Matsubara frequency index.
The two model medium responses are quite similar. Examination of the nonretarded variant of the van der Waals interaction (Eq. \ref{eq:cyl-paral-nonretarded}), 
suggests that there exists a possibility of observing interesting effects in the longitudinal ''channel'' of dielectric response. Namely, 
the product $\Delta_{1,\parallel} \Delta_{2,\parallel}$ is positive when $n<10$, negative when $10<n<200$, and again positive when 
$n>200$. Thus, the contributions to the total (summed) van der Waals interaction are both positive and negative, depending on the value of $n$. As 
retardation screens the contribution from larger values of $n$ at large separation distances, one may {\em a priori} expect to see 
a change in character of van der Waals interaction (repulsive vs. attractive) depending on the separation distance. This is indeed confirmed by 
numerical results shown in Fig. \ref{fig:fig5}.
However, the overall behavior of the van der Waals interaction, being attractive for model 1 medium, and repulsive at short distances for model 2 medium, is even more striking. This change can occur by quite a minor alteration of the medium response. Note, however, that the absolute magnitude of [5,1] - [29,0] van der Waals interaction is a factor of 200 smaller with respect to the case when the SWCNTs are in vacuum ($\sim 0.005 \; k_BT$/nm in medium vs. $\sim 1.1 \; k_BT$/nm in vacuum at $\ell = 1.32$ nm). An expected minimum in the van der Waals interaction is observed in model 2 of the medium at $\ell \approx 17$ nm, but its depth is quite small ($\sim 10^{-8} k_B T$/nm). The origin of this effect is exactly the same as in the case studied by Elbaum and Schick - the medium dielectric response (water in their case) is such that the product $[\epsilon^c_{1,\parallel}(i \omega)-\epsilon_m(i \omega)] [\epsilon^c_{2,\parallel}(i \omega)-\epsilon_m (i \omega)]$ is both positive and negative, depending on the Matsubara frequency, $i \omega_n$ (there is a crossover (or several crossovers) between the medium dielectric response and one of the cylinders' responses). In the case studied by Elbaum and Schick, the medium response (water) is quite similar to the response of one of the two half-spaces (ice). This leads us to experiment a bit more with the medium response, making it more similar to the response of one of the cylinders. The results of one such experiment are shown in Fig. \ref{fig:figmednov}.
The minimum in now located at quite small separations (3.5 nm), and its depth is larger (5.3 $10^{-6}$ $k_BT$/nm). We also observe 
a {\em maximum} ($\sim$ 2 $10^{-4}$ $k_BT$/nm at $\sim$ 1.5 nm) in a narrow repulsive region of interaction separating the two attractive interaction 
regions ($\ell < 1.4$ nm and $\ell > 2.7$ nm). This interesting effect was not observed in the previous models. We should also note here 
that the appearance of the minimum in the interaction is a quite delicate effect that needs precise tuning of the parameters of the medium 
response. From our numerical experiments we found that, while it is in general easy to obtain purely repulsive or purely attractive interactions, 
the minimum in the interaction appears as a quite special effect, and only in a narrow region of parameters describing the response of the intervening medium.

\section{Discussion}

The derivation of van der Waals - dispersion interaction between two anisotropic dielectric cylinders presented here, based on the Pitaevskii {\sl ansatz}, is valid when the interaction energy either scales linearly with the length, as in the case of parallel cylinders, or does not scale with the length at all, as in the case of skewed cylinders.  All the cases where the finite cylinder length effects come into play are not covered by this derivation.  In addition the cylinder radius has to be the smallest length involved in the problem. Another drawback of this method is that the cases of infinitely large dielectric response as in the metallic SWCNTs that show a Drude-like response at zero frequency can not be dealt with within the framework proposed here. The zero frequency term in the Matsubara summation would have to be treated differently \cite{Barash89} for those cases. 

Despite these drawbacks the present approach yields the final result for the calculation of van der Waals dispersion interactions between infinitely long semiconducting anisotropic SWCNTs. We derived all the appropriate limits and showed how the detais and peculiarities of the dielectric response of various SWCNTs effect this interaction and can in general lead also to non-monotonic van der Waals dispersion interactions of a type analogous to those found by Elbaum and Shick in the case of an aqueous layer on the surface of ice \cite{Elbaum}.

We analyzed in detail also the effect of retardation at sufficiently large inter cylinder spacings. The general conclusion is that these effects kick in at separation of about 100 nm for parallel cylinders and at about somewhat smaller separation of 50 nm for skewed cylinders. In its retarded form, that decays one distance power faster than the non-retarded form, the dispersion interactions between cylinders are in general small, only a fraction of $k_BT$ per nm length of the cylinder. Nevertheless these effects are real and can be significant for sufficiently long carbon nanotubes.

Since the dispersion interactions depend in a complicated way on the dielectric properties of the interacting anisotropic cylinders there exists a possibility that between two different types of SWCNTs, with sufficiently different dielectric spectra, the interaction would become repulsive, or even show a non-monotonic separation dependence. Such variation of van der Waals - dispersion interaction has been observed in other contexts \cite{Elbaum}. These delicate effects in themselves do not depend on the fact that the interacting cylinders are anisotropic. In fact, they persist even when the cylinders are made of isotropic dielectric materials. 

\section{Acknowledgement}

A.\v{S}. and R.P. would  like to acknowledge partial financial support for this work by 
the European Commission under Contract No. NMP3-CT-2005-013862 (INCEMS) and by the 
Slovenian Research Agency under Contract  No.  J1-0908 (Active media nanoactuators with 
dispersion forces). A.\v{S}. also acknowledges support by Croatian Ministry of Science 
(project No. 035-0352828-2837).

R.R. would like to acknowledge financial support for this 
work by the NSF grant under Contract No. CMS-0609050 
(NIRT) and the Dupont-MIT Alliance (DMA).  W.Y.C. is supported by DOE under 
Grant No. DE-FG02-84DR45170. This study was supported by the Intramural Research Program of the NIH, Eunice Kennedy Shriver National  Institute of Child Health and Human Development.

\end{document}